\documentclass[journal = jacsat]{achemso}

\usepackage{hyperref}
\usepackage{pdfpages}

\author{Benjamin Lowe}
\affiliation
{School of Physics and Astronomy, Monash University, Clayton, Victoria 3800, Australia}
\alsoaffiliation
{ARC Centre for Excellence in Low-Energy Electronics Technologies, Monash University, Clayton, Victoria 3800, Australia}
\altaffiliation{Contributed equally to this work}
\author{Jack Hellerstedt}\affiliation
{School of Physics and Astronomy, Monash University, Clayton, Victoria 3800, Australia}
\alsoaffiliation
{ARC Centre for Excellence in Low-Energy Electronics Technologies, Monash University, Clayton, Victoria 3800, Australia}
\altaffiliation{Contributed equally to this work}
\author{Adam Mat\v{e}j}
\affiliation{Institute of Physics, Academy of Sciences of the Czech Republic, Cukrovarnická 10, 1862 53, Prague, Czech Republic}
\alsoaffiliation
{Regional Centre of Advanced Technologies and Materials, Czech Advanced Technology and Research Institute (CATRIN), Palacký University Olomouc, 779 00 Olomouc, Czech Republic}
\alsoaffiliation
{Department of Physical Chemistry, Faculty of Science, Palacký University Olomouc, 771 46 Olomouc, Czech Republic}
\altaffiliation{Contributed equally to this work}
\author{Pingo Mutombo}
\affiliation{Institute of Physics, Academy of Sciences of the Czech Republic, Cukrovarnická 10, 1862 53, Prague, Czech Republic}
\author{Dhaneesh Kumar}
\affiliation{School of Physics and Astronomy, Monash University, Clayton, Victoria 3800, Australia}
\alsoaffiliation
{ARC Centre for Excellence in Low-Energy Electronics Technologies, Monash University, Clayton, Victoria 3800, Australia}
\author{Martin Ondr\'{a}\v{c}ek}
\affiliation{Institute of Physics, Academy of Sciences of the Czech Republic, Cukrovarnická 10, 1862 53, Prague, Czech Republic}
\author{Pavel Jelinek}
\affiliation{Institute of Physics, Academy of Sciences of the Czech Republic, Cukrovarnická 10, 1862 53, Prague, Czech Republic}
\alsoaffiliation
{Regional Centre of Advanced Technologies and Materials, Czech Advanced Technology and Research Institute (CATRIN), Palacký University Olomouc, 779 00 Olomouc, Czech Republic}
\email{jelinekp@fzu.cz}
\author{Agustin Schiffrin}
\affiliation{School of Physics and Astronomy, Monash University, Clayton, Victoria 3800, Australia}
\alsoaffiliation
{ARC Centre for Excellence in Low-Energy Electronics Technologies, Monash University, Clayton, Victoria 3800, Australia}
\email{agustin.schiffrin@monash.edu}

\title{Selective Activation of Aromatic C-H Bonds Catalyzed by Single Gold Atoms at Room Temperature}

\begin{document}

\begin{abstract}
Selective activation and controlled functionalization of C-H bonds in organic molecules is one of the most desirable processes in synthetic chemistry. Despite progress in  heterogeneous catalysis using metal surfaces, this goal remains challenging due to the stability of C-H bonds and their ubiquity in precursor molecules, hampering regioselectivity. Here, we examine the interaction between 9,10-dicyanoanthracene (DCA) molecules and Au adatoms on an Ag(111) surface at room temperature (RT). Characterization via low-temperature scanning tunneling microscopy, spectroscopy, and noncontact atomic force microscopy, supported by theoretical calculations,  revealed the formation of organometallic DCA-Au-DCA dimers, where C atoms at the ends of the anthracene moieties are bonded covalently to single Au atoms. The formation of this organometallic compound is initiated by a regioselective cleaving of C-H bonds at RT. Hybrid quantum mechanics/molecular mechanics calculations show that this regioselective C-H bond cleaving is enabled by an intermediate metal-organic complex which significantly reduces the dissociation barrier of a specific C-H bond. Harnessing the catalytic activity of single metal atoms, this regioselective on-surface C-H activation reaction at RT offers promising routes for future synthesis of functional organic and organometallic materials.
\end{abstract}

\section{Introduction}
The selective activation and controlled functionalization of C-H bonds in molecular precursors is one of the most important areas of synthetic organic chemistry for the design of C-C and C-heteroatom bonds in new compounds.\cite{arndtsen_selective_1995,labinger_understanding_2002,godula_c-h_2006} Controlled C-H activation reactions have enticing applications including natural gas transport,\cite{labinger_understanding_2002,crabtree_alkane_2001,schwach_direct_2017} polymer fabrication,\cite{blaskovits_c-h_2019,yang_metal-catalyzed_2018,zhang_recent_2018} and late-stage modification of pharmaceutical products.\cite{dai_divergent_2011,jana_emergence_2021,wencel-delord_ch_2013,zhang_perspective_2022,cernak_medicinal_2016,yamaguchi_c-h_2012} The high bond-dissociation enthalpy and ubiquity of C-H bonds, however, make their selective cleavage challenging.\cite{arndtsen_selective_1995,dutta_arene_2021,ramadoss_remote_2022,xue_essential_2017} In solution, enzymatic and transition metal (homogeneous) catalysis have provided avenues for the selective activation of C-H bonds.\cite{yang_metal-catalyzed_2018,tiwari_catalyst-controlled_2019,shilov_activation_1997,labinger_platinum-catalyzed_2017,lewis_enzymatic_2011}

Solid metal surfaces have also been shown to exhibit (heterogeneous) catalytic activity on adsorbed molecules.\cite{clair_controlling_2019,fan_surface-catalyzed_2015,grill_covalent_2020,held_covalent-bond_2017} In particular, controlled activation of aromatic C-H bonds on a surface has enabled the synthesis of low-dimensional organic nanomaterials with promising electronic and magnetic properties, such as nanographenes,\cite{mishra_topological_2020,mishra_collective_2020,pavlicek_synthesis_2017,su_atomically_2019} graphene nanoribbons,\cite{cai_atomically_2010,rizzo_topological_2018,ruffieux_-surface_2016} organic polymers,\cite{cirera_tailoring_2020,sanchezgrande_-surface_2019} and covalent organic frameworks\cite{galeotti_synthesis_2020,springer_topological_2020}. However, broader application of on-surface C-H activation for the synthesis of functional materials remains limited due to high-bond energies and poor regioselectivity.\cite{fan_precise_2018,li_surface-controlled_2016}

A common reaction used for on-surface design of aromatic C-C and C-heteroatom bonds is Ullmann-type coupling, where C-X bonds (X: halogen atom) in precursor molecules undergo surface-catalyzed dehalogenation.\cite{lackinger_surface-assisted_2017} This approach has been successfully exploited for the fabrication of a range of different materials,\cite{clair_controlling_2019,fan_surface-catalyzed_2015,grill_covalent_2020,held_covalent-bond_2017,lackinger_surface-assisted_2017} although the requirement for pre-halogenation of targeted carbons in precursor molecules limits its applicability.

One promising approach for on-surface aromatic C-H activation consists of leveraging the catalytic activity of single metal adatoms. Single-atom catalysis yields promise for reducing the amount of precious metals used in heterogeneous catalysis.\cite{giannakakis_single-atom_2019,mitchell_single_2020,parkinson_single-atom_2019,wang_heterogeneous_2018} In particular, despite its relative chemical inertness, gold has emerged as a promising single-atom catalyst (e.g., for the oxidation of carbon monoxide).\cite{parkinson_single-atom_2019,wang_heterogeneous_2018,bohme_gas-phase_2005} While Au surfaces have been used to catalyze the cleavage of C-H and C-X bonds,\cite{zhong_linear_2011,li_surface-controlled_2016,lafferentz_controlling_2012,blunt_templating_2010} sometimes resulting in the formation of organometallic C-Au bonds,\cite{zhang_surface_2014,xing_selective_2019,karan_gold-linked_2020} atomic-scale studies of single Au atom catalysis for C-H activation (on surfaces other than Au) have not yet been conducted.

Herein, we report the regioselective cleavage of an aromatic C-H bond of a cyano-functionalized anthracene derivative, 9,10-dicyanoanthracene (DCA; Figure \ref{1}a), activated by the interaction with single Au atoms on a Ag(111) surface at room temperature (RT). This selective activation of a specific C-H bond (located at the end of the anthracene group - labelled position A in Figure \ref{1}a) is mediated by the on-surface formation of an intermediate DCA-Au metal-organic complex, which subsequently enables selective dehydrogenation of the DCA molecule (Figure \ref{1}d). This results in a reactive site at the DCA anthracene extremity, which can then covalently bond to a Au adatom and form organometallic DCA-Au-DCA dimers via a C-Au-C motif (Figure \ref{1}e). We used low-temperature scanning tunneling microscopy (STM) and spectroscopy (STS) as well as non-contact atomic force microscopy (ncAFM) to characterize the organometallic dimers at the atomic scale. The experimental findings are supported by density functional theory (DFT), ncAFM simulations and quantum mechanics/molecular mechanics (QM/MM) calculations to shed light on the atomic structure and reaction mechanism. The low activation energy, indicated by the reaction occurring at RT, and regioselectivity of the observed C-H activation open the door to unexplored on-surface synthesis protocols based on single atom catalysis for the development of functional and robust organic and organometallic materials. 
\section{Results}

\begin{figure}[H]
    \centering
    \includegraphics{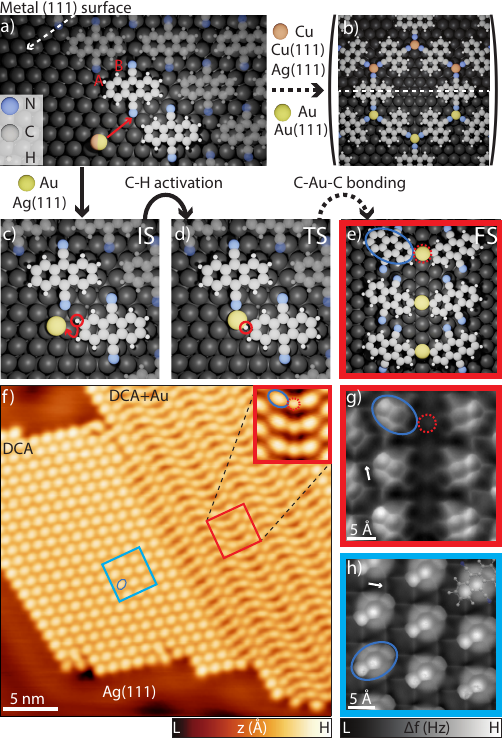}
    \caption{Interactions between noble metal adatoms and cyano-functionalized aromatic molecules on noble metal (111) surfaces: metal-cyano coordination vs. covalent metal-carbon bonding. a) Schematic of 9,10-dicyanoanthracene (DCA) molecules and metal adatoms (Cu or Au) deposited on a (111) noble metal surface (Ag, Cu or Au), with C-H positions A (anthracene end) and B (cyano adjacent) labelled. b) The interaction between DCA and Cu adatoms on Ag(111) or on Cu(111), or with Au adatoms on Au(111), leads to formation of a 2D honeycomb-kagome metal-organic framework (MOF) via three-fold rotationally symmetric metal-cyano coordination\cite{pawin_surface_2008,zhang_probing_2014,hernandez-lopez_searching_2021,kumar_manifestation_2021,yan_-surface_2019}. c)-e) The interaction between DCA and Au adatoms on Ag(111) gives rise to organometallic DCA-Au-DCA dimers via selective (position A) C-H  activation and subsequent C-Au covalent bonding (IS: initial state, TS: transition state, FS: final state). f) Constant-current STM image of self-assembled organic DCA-only (DCA) and DCA+Au domains featuring organometallic DCA-Au-DCA dimers on Ag(111) ($V_{\mathrm{b}} = -20$ mV, $I_{\mathrm{T}} = 25$ pA). Inset: high-resolution STM image (same tunneling parameters) of cis organometallic DCA-Au-DCA dimers within DCA+Au self-assembly. Solid blue ellipse: individual DCA molecule. Dashed red circle: Au adatom. g)-h) Constant-height ncAFM  images (CO-tip; tip 30 pm closer to the sample with respect to STM setpoint $V_{\mathrm{b}} = 15$ mV, $I_{\mathrm{t}} = 100$ pA) of DCA domain (blue box in f) and of three cis DCA-Au-DCA organometallic dimers in DCA+Au domain (red box in f). White arrows indicate intermolecular N\--\--H interactions.
}
    \label{1}
\end{figure}

We co-deposited DCA molecules and Au atoms from the gas phase in ultra-high vacuum (UHV) onto a clean Ag(111) surface held at RT (see Methods for further details). Figure \ref{1}f shows an STM image of a sample region following this preparation. We observe two types of well-ordered, two-dimensional (2D) periodic arrays of self-assembled adsorbates: (i) a purely organic domain (cyan box, region labelled DCA) consisting of identical elliptical protrusions (solid blue outline) corresponding to non-covalently bonded DCA molecules; \cite{kumar_electric_2019} (ii) a domain consisting of parallel rows of pairs (red box, region labelled DCA+Au) of similar elliptical features (inset: solid blue outline) linked by circular protrusions (inset: red dashed circular outline). Based on their appearance and size, we identify the elliptical features in the DCA+Au domain as DCA molecules,\cite{kumar_electric_2019} arranged in pairs in either cis (red box in Figure \ref{1}f) or trans configuration (see SI Figure S1). Approximately equal occurrences of cis- and trans- configuration were observed, though a quantitative statistical analysis is beyond the scope of this study.

The qualitative difference in Figure \ref{1}f between the previously reported\cite{kumar_electric_2019} DCA domain and the DCA+Au domain consisting of ordered DCA pairs is a strong suggestion that the pairing of DCA molecules in the latter is mediated by the interaction between DCA and Au adatoms. Notably, these results are also qualitatively different from previously studied surface-supported systems where 3-fold rotationally symmetric metal-ligand coordination between DCA cyano groups and noble metal adatoms [e.g., Cu adatoms on Cu(111),\cite{pawin_surface_2008,zhang_probing_2014,hernandez-lopez_searching_2021} Ag(111),\cite{kumar_manifestation_2021} Ir(111)-supported graphene,\cite{yan_synthesis_2021} and NbSe\textsubscript{2}\cite{yan_two-dimensional_2021}; Au adatoms on Au(111)\cite{yan_-surface_2019}] give rise to porous, 2D honeycomb-kagome metal-organic frameworks (MOFs) as shown in Figure \ref{1}b. This is a strong indication that the interaction between DCA and Au adatoms on Ag(111) is fundamentally different than in these other systems.

We used ncAFM with a carbon-monoxide (CO)-functionalized tip (see Methods) to gain further insight into the intramolecular structure of the Au-mediated DCA pairs. NcAFM imaging of the DCA+Au domain, Figure \ref{1}g, and the DCA domain,\cite{kumar_electric_2019} Figure \ref{1}h, are similar in that the submolecular structure of individual DCA molecules is clearly resolved in agreement with the overlaid chemical structure in Figure \ref{1}h, with non-covalent attractive N\--\--H interactions (Figure 1 white arrows) between adjacent molecules (DCA) or molecule pairs (DCA+Au).\cite{arras_nature_2012}

In contrast with ncAFM imaging of the DCA domain in Figure \ref{1}h where the frequency shift ($\Delta f$) across each of the DCA benzene rings is relatively uniform, individual DCA molecules in the DCA+Au domain, Figure \ref{1}g, exhibit reduced $\Delta f$ at the benzene rings closest to the DCA pair center (near the location of the circular protrusion in STM imaging; red dashed circle). Such ncAFM appearance of $\pi$-conjugated rings is indicative of direct coordination between atoms of the aromatic group and a metal adatom,\cite{zhang_coordination-controlled_2019,pawlak_bottom-up_2020,li_hierarchical_2018,fan_chemical_2020} further supporting the involvement of Au adatoms within the DCA+Au domain.

\begin{figure}[H]
    \centering
    \includegraphics{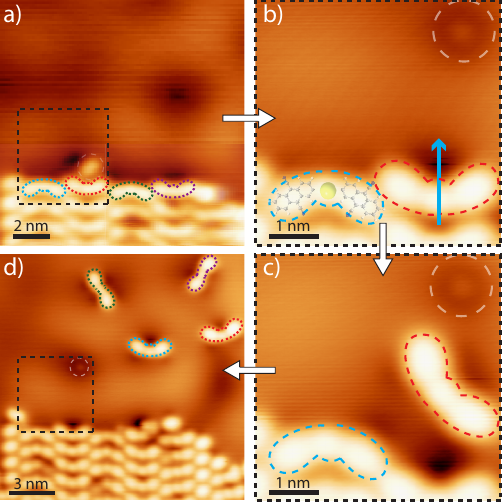}
    \caption{Deconstruction of organometallic DCA+Au self-assembly on Ag(111) via lateral STM manipulation. a)-d) Constant-current STM images showing manipulation of single DCA-Au-DCA units from cis DCA+Au domain [images: $V_{\mathrm{b}} = -20$ mV, $I_{\mathrm{t}} = 10$ pA; manipulation: $V_{\mathrm{b}} = -10$ mV, $I_{\mathrm{t}} = 14$ nA for removal of DCA dimers from 2D film, $V_{\mathrm{b}} = -10$ mV, $I_{\mathrm{t}} = 5$ nA for subsequent translation on Ag(111)]. Blue arrow indicates STM tip path during manipulation. Dashed black square in a) and d) indicates region in b) and c). Dashed outlines indicate manipulated DCA-Au-DCA units. White dashed circles indicate defect adsorbate displaced during manipulation. Ball-and-stick model of DCA-Au-DCA dimer chemical structure overlaid in b).}
    \label{2}
\end{figure}

We conducted lateral STM manipulation to further probe the nature of the intra- and inter-DCA pair interactions within the DCA+Au assembly. Figure \ref{2}a shows an STM image of a DCA+Au domain with cis-configuration DCA pairs. Individual DCA pairs, consisting of two elliptical features connected via a circular feature in STM images, could be reproducibly removed from the edge of a DCA+Au domain and manipulated about the surface (Figure \ref{2}b-d). Upon manipulation, the STM appearance of the DCA pairs is unchanged, suggesting that their chemical structure is preserved.

In some instances, we observed switching from the cis- to the trans- configuration upon manipulation (dashed green and purple pairs in Figure \ref{2}). We did not, however, observe evidence of such switching under normal tunneling conditions (through e.g. telegraph noise\cite{kumagai_controlling_2014}); cis- and trans- configurations were both otherwise stable during characterization.

Within the range of STM manipulation parameters tested, we were unable to observe further breaking of the DCA pairs into their individual constituents (i.e. STM elliptical features corresponding to DCA molecules and circular features). From this, we conclude that while lateral manipulation can overcome the strength of bonding between adjacent DCA pairs, it cannot overcome the interaction connecting the two DCA molecules within such pairs. This suggests the intra-pair interaction is significantly stronger than the inter-pair interaction, and that the Au-mediated DCA pairs are highly stable.

\begin{figure}[H]
    \centering
    \includegraphics{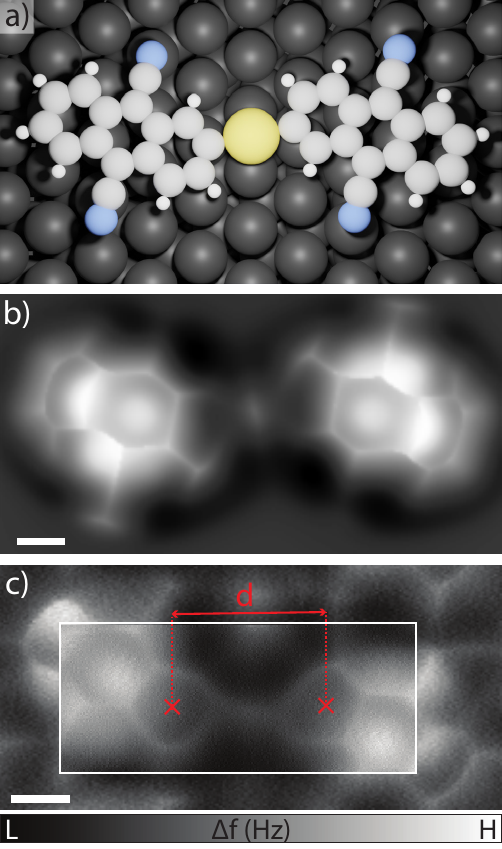}
    \caption{Experimental and theoretical submolecular structure of a trans DCA-Au-DCA pair on Ag(111). a) Ball-and-stick model of DFT-relaxed structure. b) Simulated ncAFM image of optimized structure in a). c) Constant-height  (CO-tip) ncAFM image (tip 30 pm closer to the sample with respect to STM setpoint $V_{\mathrm{b}} = 15$ mV, $I_{\mathrm{t}} = 100$ pA). Inset in white box is a higher resolution image taken with the tip 40 pm closer to the surface with respect to the same STM setpoint (see Methods). Red arrow indicates distance, $d$, between the centers of the benzene rings closest to the Au atom of each DCA molecule involved in an organometallic pair. Simulated image in b): $d_{\textrm{sim}}= 5.72 \pm 0.06$ Å. Experimental image in c): $d_{\textrm{exp}}=6.0 \pm 0.2$ Å. Scale bars: 2.5 \r{A}.
}
    \label{3}
\end{figure}

From these STM manipulation observations, combined with the reduced frequency shift of the benzene rings near the DCA pair center (Figure \ref{1}f, \ref{3}c), we propose that the DCA pairs consist of an organometallic DCA-Au-DCA motif, wherein a position A (Figure \ref{1}a) carbon atom of each of the two DCA molecules is covalently bonded to a Au atom (Figure \ref{1}e). The observation of cis- and trans- configurations can be attributed to the chemical equivalence of the two carbon atoms at the anthracene ends of the DCA molecule.

We performed DFT calculations to rationalize the proposed organometallic structure. The energetically favorable atomic structure of an organometallic trans DCA-Au-DCA dimer on Ag(111), optimized via DFT (see Methods), is shown in Figure \ref{3}a). The DFT calculations revealed a negligible energy difference between the trans and cis configurations (see SI Figure S2), supporting the observation of the two configurations in approximately equal measure, and the cis-to-trans conversion by STM manipulation.

 The simulated ncAFM image (Figure \ref{3}b; obtained via the probe-particle method \cite{hapala_mechanism_2014}) corresponding to the DFT-relaxed DCA-Au-DCA dimer in Figure \ref{3}a shows strong qualitative agreement with the experimental ncAFM image (Figure \ref{3}c), including the reduced $\Delta f$ of the benzene rings closest to the C-Au-C bonding motif at the DCA pair center. The simulated and experimental image also have a similar distance $d$ between the centers of the benzene rings closest to the Au atom (5.72 $\pm$ 0.06 \r{A} for the simulated images and 6.0 $\pm$ 0.2 \r{A} for the experimental images). We rule out a DCA-DCA pair with a direct covalent C-C bond between facing anthracene groups, since ncAFM simulations of this structure would result in a distance $d$ (4.76 $\pm$ 0.06 \r{A}) significantly shorter than that observed experimentally, and would not reproduce the experimentally observed reduced $\Delta f$ in the benzene rings closest to the DCA-DCA linkage (Figure S3).

\begin{figure}[H]
    \centering
    \includegraphics{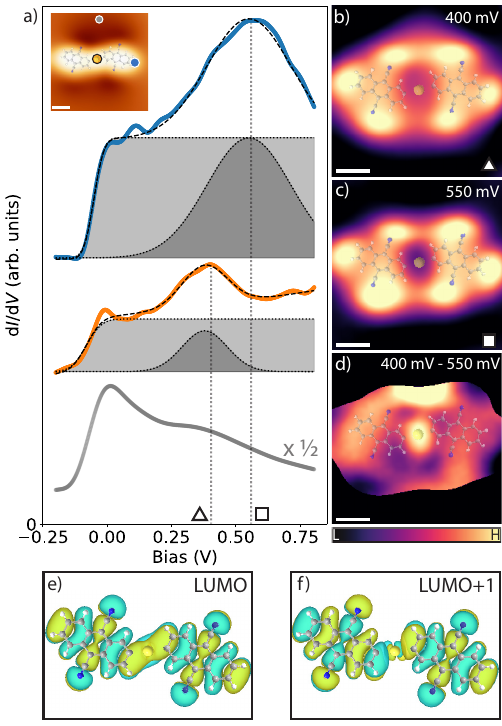}
    \caption{Electronic properties of DCA-Au-DCA organometallic unit. a) $\mathrm{d}I/\mathrm{d}V$ scanning tunneling spectra obtained at sites indicated in inset (setpoint: $V_\mathrm{b} = -200$ mV, $I_\mathrm{t} = 20$ pA). Black dashed curves: fits of experimental data consisting of an error function (step-like feature, light gray shaded area), a Gaussian peak (dark gray). An exponential function accounting for tunneling transmission was also included in the fit for the orange curve. Silver reference spectrum (gray curve) scaled down by a factor of 2. Vertical dotted gray lines indicate bias voltages at which $\mathrm{d}I/\mathrm{d}V$ maps in (b) and (c) were acquired. Inset: constant-current STM image of DCA-Au-DCA unit ($V_\mathrm{b} = -20$ mV, $I_\mathrm{t} = 50$ pA).  b), c) Constant-current $\mathrm{d}I/\mathrm{d}V$ maps of DCA-Au-DCA unit ($I_\mathrm{t} = 500$ pA) at $V_\mathrm{b} = 400$ and $550$ mV. d) Subtraction between $\mathrm{d}I/\mathrm{d}V$ maps in b), c) for region corresponding to DCA-Au-DCA unit. DCA-Au-DCA chemical structure overlaid for reference in a) inset and b)-d). Scale bars: 5 \r{A}. We isolated the single DCA-Au-DCA unit from a DCA+Au self-assembled 2D organometallic domain via lateral STM manipulation. e)-f) Wavefunction isosurfaces (0.005 \r{A}$^{-3/2}$; green: positive; blue: negative) for LUMO and LUMO+1 of gas-phase trans DCA-Au-DCA pair calculated by DFT.  
}
    \label{4}
\end{figure}

We performed $\mathrm{d}I/\mathrm{d}V$ STS to gain insight into the electronic structure of an individual DCA-Au-DCA dimer. The spectra in Figure \ref{4}a were acquired on the Au (orange) and outer (i.e. furthest from Au site) anthracene end (blue) sites of a trans DCA-Au-DCA dimer, removed from the 2D assembly via manipulation. Both orange and blue spectra show a step-like feature at $\sim$-66 mV, and a peak at $\sim$376 and $\sim$560 mV, respectively.

We attribute the step-like feature to the Shockley surface state of Ag(111),\cite{li_local_1997} and the two peaks to intrinsic electronic states of the DCA-Au-DCA unit. Figure \ref{4}b,c show $\mathrm{d}I/\mathrm{d}V$ maps associated with each of these electronic states. Both maps show significant intensity at the anthracene ends furthest from the Au site, and at the cyano groups. These features are qualitatively similar to the spatial distribution of the DCA lowest unoccupied molecular orbital (LUMO).\cite{kumar_electric_2019,liljeroth_single-molecule_2010} The map in Figure \ref{4}b also shows some $\mathrm{d}I/\mathrm{d}V$ intensity at the Au site, significantly more so than Figure \ref{4}c (emphasized in the difference between Figure \ref{4}b and Figure \ref{4}c, displayed in Figure \ref{4}d).

Figure \ref{4}e,f shows the DFT-calculated spatial distribution of the LUMO and LUMO+1 of the structurally optimized gas-phase DCA-Au-DCA dimer (see Methods and Figure S4 in the SI). These DFT-calculated LUMO and LUMO+1 are separated in energy by $\sim$100 meV, on the same order of magnitude as the $\sim$184 meV energy difference between experimental $\mathrm{d}I/\mathrm{d}V$ peaks in Figure \ref{4}a. Both the calculated LUMO and LUMO+1 have predominantly DCA orbital character (see Figure S4 and Figure S5), with intensity at the anthracene ends furthest from the Au site, and at the cyano groups. The LUMO (Figure \ref{4}e) shows intensity at the Au atom whereas the LUMO+1 (Figure \ref{4}f) exhibits a nodal plane at this site perpendicular to the DCA-DCA axis. The calculated LUMO and LUMO+1 are in qualitative agreement with the experimental $\mathrm{d}I/\mathrm{d}V$ maps in Figure \ref{4}b and c, respectively. This agreement provides further evidence that the DCA+Au self-assembled 2D domain in Figure \ref{1}f) consists of organometallic DCA-Au-DCA dimers, in which the position A carbon atoms are covalently bonded to a Au atom.

The organometallic DCA-Au-DCA structure is qualitatively different to previously observed 2D MOFs consisting of DCA molecules coordinated with metal atoms via their cyano groups.\cite{pawin_surface_2008,zhang_probing_2014,hernandez-lopez_searching_2021,yan_synthesis_2021,kumar_manifestation_2021,yan_two-dimensional_2021,yan_-surface_2019} Moreover, the formation of the organometallic DCA-Au-DCA motif requires highly regioselective cleavage of the position A aromatic C-H bonds to enable the C-Au-C covalent bonding. Such C-H bonds have large dissociation enthalpies\cite{liu_exploiting_2022,sun_-surface_2014,zhang_coordination-controlled_2019}, $\Delta H$, meaning their cleavage seldom occurs at RT,\cite{liu_bromine_2018} and they often suffer from poor regioselectivity.\cite{zhang_coordination-controlled_2019,sun_-surface_2014,liu_exploiting_2022,li_surface-controlled_2016,held_covalent-bond_2017}. For a gas-phase DCA molecule for instance, we calculated $\Delta H$ values of 112.1 and 113.2 kcal/mol (4.86 and 4.91 eV) via DFT for the aromatic C-H bonds at positions A and B, respectively (see Figure \ref{1}a and SI Figure S7).  

In the following, we rationalize the highly selective C-H activation reaction (and subsequent DCA-Au-DCA formation) at RT via DFT and QM/MM calculations. First, we found using DFT  that Au adatoms on Ag(111) retain their neutral, atomic-like character (SI Figure S6). This contrasts with the case of Au adatoms on Au(111), where significant electronic hybridization between adatom and the surface takes place. That is, Au adatoms are less coordinated and arguably more reactive on Ag(111) than on Au(111), providing an explanation for the differences between organometallic (covalent C-Au bonding) and metal-organic\cite{yan_-surface_2019} (N\--\--Au) coordination complexes (Figure \ref{1}).

\begin{figure}[H]
    \centering
    \includegraphics{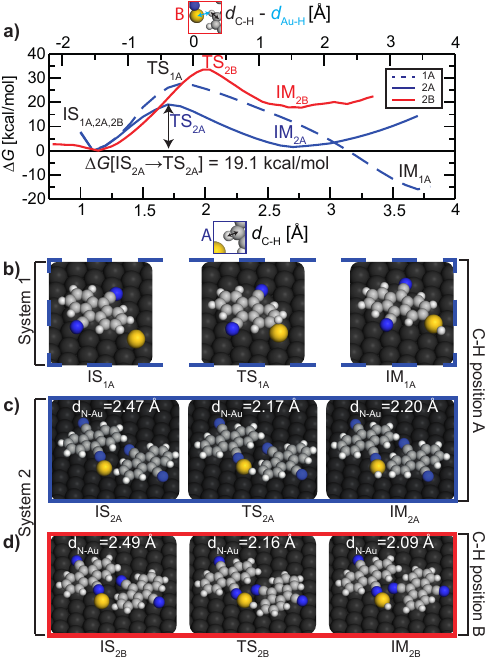}
    \caption{Pathways for Au-induced C-H cleavage of DCA molecule on Ag(111). a) Gibbs free energy differences, $\Delta G$, calculated via QM/MM simulations at $T$ = 300 K, for different C-H bond dissociation pathways in a system comprised of a single DCA molecule and a single Au adatom (system 1 in b; dashed curve), and in a system comprised of 2 DCA molecules and a single Au adatom (system 2 in c,d; solid curves). For position A (blue curves), we considered C-H bond length $d_\mathrm{{C-H}}$ (bottom axis) as reaction coordinate; for position B (red curve), we considered the difference $(d_\mathrm{{C-H}}-d_\mathrm{{Au-H}})$ (top axis) between C-H bond length and Au-H distance as reaction coordinate (computationally necessary for this pathway). Activation energy [$\Delta G$ between transition (TS) and initial (IS) states] for C-H bond dissociation is reduced for position A in system 2, $\Delta G[\mathrm{IS}_{\mathrm{2A}} \rightarrow \mathrm{TS}_{\mathrm{2A}}]=19.1$ kcal/mol (0.83 eV). b-d) Ball-and-stick models of IS, TS and intermediate (IM) states for C-H bond (positions A and B) dissociation pathways in systems 1 (b) and 2 (c, d).}
    \label{5}
\end{figure}

We then conducted QM/MM calculations of Gibbs free energy differences, $\Delta G$, on a system composed of a single DCA molecule with a single Au adatom in proximity on Ag(111) (system 1; see Figures \ref{5}b and S7). We considered a C-H bond dissociation mechanism mediated by the direct interaction between Au and H at $T = 300$ K ($\sim$RT to match our experimental conditions; see Figures S7 and S8 for calculations at $T=0$ K). The dashed blue curve in Figure \ref{5}a) shows the calculated $\Delta G$ as a function of position A C-H distance, $d_{\mathrm{C-H}}$, when system 1 evolves from an initial state IS$_{\mathrm{1A}}$ to a transition state TS$_{\mathrm{1A}}$ then to an intermediate state IM$_{\mathrm{1A}}$ (where H  becomes bonded to Au; Figure \ref{5}b). These calculations yielded an activation energy $\Delta G\left[\mathrm{IS}_{\mathrm{1A}}\rightarrow \mathrm{TS}_{\mathrm{1A}}\right]=$ 27.7 kcal/mol (1.20 eV). We claim that this energy is too high to explain the dissociation of an anthracene C-H bond at RT. Note that similar calculations performed for the position B C-H bond in system 1 suggest an even higher dissociation barrier (Figure S7). 

Therefore, we also considered a system composed of two DCA molecules and a single Au adatom on Ag(111) (system 2; Figures 5c and S8), with a configuration similar to that at the edges of self-assembled DCA-only domains (Figures \ref{1}a and S14).\cite{kumar_electric_2019} We suggest that Au atoms diffusing on the surface may encounter supramolecular DCA islands rather than isolated molecules (Figure S14 and S15). This is supported by the fact that, experimentally, we observed DCA-Au-DCA dimers both when DCA and Au were deposited sequentially (DCA first) or simultaneously (see Methods). Furthermore, we did not observe any DCA+Au domains which were surrounded by DCA-only domains. Based on this, we considered the possibility that N\--\--Au coordination could give rise to a metal-organic DCA(N)\--\--Au complex (see initial states IS$_{\mathrm{2A}}$ or IS$_{\mathrm{2B}}$ in Figure \ref{5}c,d) at the edges of DCA-only domains, and conducted QM/MM simulations to investigate whether this geometry could facilitate C-H activation at an adjacent molecule. The solid blue curve in Figure \ref{5}a) shows $\Delta G$($T$ = 300 K) as a function of position A C-H distance, $d_{\mathrm{C-H}}$, for system 2 as it evolves from IS$_\mathrm{2A}$ to an intermediate state IM$_{\mathrm{2A}}$ via a transition state TS$_{\mathrm{2A}}$ (Figure \ref{5}c). Here, we found a position A C-H activation energy $\Delta G\left[\mathrm{IS}_{\mathrm{2A}}\rightarrow \mathrm{TS}_{\mathrm{2A}}\right]=$ 19.1 kcal/mol (0.83 eV), significantly smaller than $\Delta G\left[\mathrm{IS}_{\mathrm{1A}}\rightarrow \mathrm{TS}_{\mathrm{1A}}\right]$. 

Similarly, the red curve in Figure \ref{5}a) corresponds to $\Delta G$($T$ = 300 K) as a function of $d_{\mathrm{C-H}} - d_{\mathrm{Au-H}}$ for the position B C-H bond, for the same system 2, evolving from IS$_{\mathrm{2B}}$ to TS$_{\mathrm{2B}}$ to IM$_{\mathrm{2B}}$ (Figure \ref{5}d). Here, the activation barrier $\Delta G\left[\mathrm{IS}_{\mathrm{2B}}\rightarrow \mathrm{TS}_{\mathrm{2B}}\right]$ for the position B C-H bond is 33.6 kcal/mol (1.46 eV), significantly larger than $\Delta G\left[\mathrm{IS}_{\mathrm{2A}}\rightarrow \mathrm{TS}_{\mathrm{2A}}\right]$. Based on these QM/MM calculations, the dissociation barrier of the position A C-H bonds is significantly lowered via interaction with a DCA(N)\--\--Au metal-organic complex, enabling highly regioselective C-H bond cleavage at RT.

\section{Discussion}
To gain more insight into the regioselective lowering of the activation energy of the C-H cleavage, we analyzed the electronic structure of the DCA(N)\--\--Au complex (present in IS$_\mathrm{2A}$) using DFT (Figure S10b). These calculations indicate a weak, purely electrostatic interaction between Au and cyano group, with no evidence of strong hybridization, and with an interaction energy of 4.2 kcal/mol (0.18 eV) and a N\--\--Au distance of 2.47 \r{A}. This interaction with the cyano group results in a local polarization around and charge redistribution within the Au atom, with an electron depletion on the outside of the Au atom along the Au-cyano axis (Figure S10d). This depleted electron density can reduce the Pauli repulsion between Au and a H atom of an adjacent DCA, facilitating the formation of a Au-H chemical bond and of a DCA(H)-Au\--\--(N)DCA complex (see IM$_\mathrm{2A}$ in Figure \ref{5}c). This partially explains the significant lowering of the C-H bond activation energy. 

Our DFT calculations also show that, in IM$_{\mathrm{2A}}$ (Figure \ref{5}), covalent Au-H bonding is concomitant with hybridization of Au 6s and 5d orbitals, giving rise to two hybrid sd orbitals along the N\--\--Au-H axis (see SI Section S9, Figures S9-S13), and with electron transfer from Au to H (Figure S12). The resulting partial positive charge on Au strengthens the electrostatic interaction with the partially negatively charged cyano group (Figure S11), causing an increase in the Au\--\--(N)DCA interaction energy of $\sim$22 kcal/mol ($\sim$0.95 eV) and a reduction of the Au\--\--N bond length by $\sim$0.27 \r{A}. 

To understand how this change in Au\--\--(N)DCA interaction energy influences the C-H cleavage activation barrier, it is pertinent to consider the transition state TS$_{\mathrm{2A}}$ (Figure \ref{5}). Here, DFT calculations show a Au\--\--(N)DCA interaction energy of 26.0 kcal/mol (1.13 eV) with a Au\--\--N bond length of 2.17 \r{A} - indicating a much stronger interaction than in IS$_{\mathrm{2A}}$. This synergistic effect therefore contributes further to the significant lowering of the activation energy of the C-H cleavage at position A. 

Moreover, we claim that, for the position B C-H bond (Figure \ref{5}d), this effect is offset by steric hindrance and N-N repulsion between adjacent DCA molecules, contributing to a significantly larger dissociation barrier $\Delta G\left[\mathrm{IS}_{\mathrm{2B}}\rightarrow \mathrm{TS}_{\mathrm{2B}}\right]$. 

For TS$_{\mathrm{2A}}$, we calculated via DFT a C-H stretching mode in the direction of the C-H dissociation with an eigenfrequency of 611.5 cm$^{-1}$. Using this value, we calculated for position A (with a dissociation barrier of 19 kcal/mol) at RT an Arrhenius \cite{telychko_strain-induced_2019} C-H dissociation rate constant of 0.23 s$^{-1}$. By contrast, the RT Arrhenius dissociation rate constant for the position B C-H bond (with a dissociation barrier of 33.6 kcal/mol) is $1.4\times 10^{-12}$ s$^{-1}$. We claim that this significant difference between positions A and B explains the experimentally observed regioselectivity of the C-H dissociation. The 19 kcal/mol barrier represents a significant reduction relative to comparable C-H activation reactions with barriers of 30-40 kcal/mol (1.3-1.8 eV), experimentally occurring in a temperature window of 400-700 K.\cite{zhang_coordination-controlled_2019,liu_exploiting_2022,sun_-surface_2014,han_sequential_2022} Furthermore, other on-surface reactions with barriers of $\sim$20 kcal/mol ($\sim$0.87 eV) have been observed at RT,\cite{telychko_strain-induced_2019,zeng_chemisorption-induced_2022} supporting the interpretation of our experimental results as RT C-H activation.

Although our DFT and QM/MM calculations rationalize the regioselective cleaving of position A C-H bonds at RT, they do not explain the pathway from IM$_{\mathrm{2A}}$ (Figure \ref{5}) to the experimentally observed organometallic DCA-Au-DCA motif (where C bonds covalently to Au). In the following, we provide a tentative, qualitative scenario for such a pathway. We propose that, at RT, the hydrogen atom in DCA(N)\--\--Au-H in IM$_{\mathrm{2A}}$ (Figure \ref{5}c) is likely to detach, bind to Ag(111) and then desorb (see SI Section S10), leaving organometallic DCA(C)\--Au on Ag(111) which can diffuse on the surface and potentially react with another dehydrogenated DCA$\cdot$ radical, forming the experimentally observed DCA-Au-DCA unit. Non-covalent N\--\--H bonds between neighboring DCA-Au-DCA dimers may contribute to stabilizing the DCA-Au-DCA dimer rows as seen in the DCA-Au domain in Figure \ref{1}f. The relatively weak and reversible DCA(N)\--\--Au interaction, combined with the reactivity of the DCA$\cdot$ radical, might explain why we did not experimentally observe any of the metal-organic DCA(N)\--\--Au species such as in IS$_\mathrm{2A}$ or IM$_\mathrm{2A}$ in Figure \ref{5}. A quantitative theoretical description of this process would require the consideration of many possible reaction pathways (including, for example, surface diffusion, dynamic equilibrium of different phases) in a system composed of many DCA molecules, Au adatoms and DCA-Au complexes; this is beyond the scope of this work.

At $T=$ 0 K, QM/MM calculations suggest that the energies involved in the dissociation of a position A C-H bond in a DCA-Au-DCA dimer (by interacting with a DCA(N)\--\--\--Au-H complex; see SI Figure S16) are approximately twice as large as that for a position A C-H bond in a single DCA molecule. This explains the observation of discrete DCA-Au-DCA dimers and absence of 1D -[DCA-Au-DCA-Au]- organometallic chains.

We claim that the key difference between the previously observed DCA-Cu honeycomb-kagome MOF on Ag(111),\cite{kumar_manifestation_2021} and the present study on the organometallic DCA-Au motif is the strength of the metal adatom-cyano interaction. For Cu\--\--N, DFT calculations revealed an interaction energy of $\sim$18.5 kcal/mol ($\sim$0.8 eV), significantly larger than the aforementioned value of 4.2 kcal/mol for Au\--\--N in IS$_{\mathrm{2A}}$ (Figure \ref{5}c). This favors three-fold Cu-ligand coordination and Cu-based MOF formation, preventing the reaction pathway for C-H bond activation and organometallic bonding as is the case for Au.

\section{Conclusion}
We have experimentally observed the formation of organometallic DCA-Au-DCA dimers on a Ag(111) surface at RT. In these dimers, C atoms located specifically at the extremity of the DCA molecule anthracene moiety (position A) bond covalently to a Au adatom. These observations, supported by DFT, indicate the regioselective C-H bond activation triggered by Au adatoms at RT. Our QM/MM calculations show that this reaction is mediated by the interaction with an intermediate DCA(N)\--\--Au-H metal-organic complex, which significantly reduces the activation barrier of a specific C-H bond. Our work provides atomic-scale insight into selective C-H bond activation, catalyzed by single noble metal atoms and intermediate metal-organic complexes at RT. Our findings have the potential to contribute to addressing the large demand for the synthesis of fuel, fine chemicals and functional organic and organometallic materials. Further investigations could explore such protocols on surfaces or host materials beyond noble metals. 

\section{Acknowledgements}
A.S.  acknowledges   funding   support from the ARC Future Fellowship scheme (FT150100426). B.L., D.K., and J.H. acknowledge funding support from the Australian Research Council (ARC) Centre of Excellence in Future Low-Energy Electronics Technologies (CE170100039). B.L. is supported through an Australian Government Research Training Program (RTP) Scholarship. A.M. acknowledges the support from internal student grant agency of Palacky University in Olomouc (IGA\textunderscore PrF\textunderscore2022\textunderscore019). P.M, A.M., M.O. and P.J. acknowledge support of Praemium Academie of the Academy of Science of the Czech Republic and Czech Science Foundation (project no. 20-13692X). We acknowledge computational resources provided by the project ``e-Infrastruktura CZ" (e-INFRA CZ LM2018140) supported by the Ministry of Education, Youth and Sports of the Czech Republic.

\section{Supporting Information}
Experimental and theoretical methods; structural characterization of DCA+Au domains; DFT optimized structures; further ncAFM simulations; electronic properties of DCA+Au dimers; Au adatom character on Ag(111); QM/MM calculations at $T = 0$ K; hybridization and charge distribution across reaction pathway; tentative organometallic dimer formation pathway; secondary C-H activation QM/MM calculations.

\begin{tocentry}
    \includegraphics{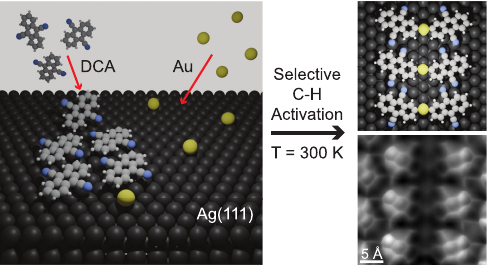}
    Deposition of Au atoms and 9,10-dicyanoanthracene (DCA) molecules on Ag(111) at room temperature (300 K) results in selective C-H activation to form organometallic DCA-Au-DCA dimers. 
\end{tocentry}
\bibliography{jacs}

\includepdf[pages=-]{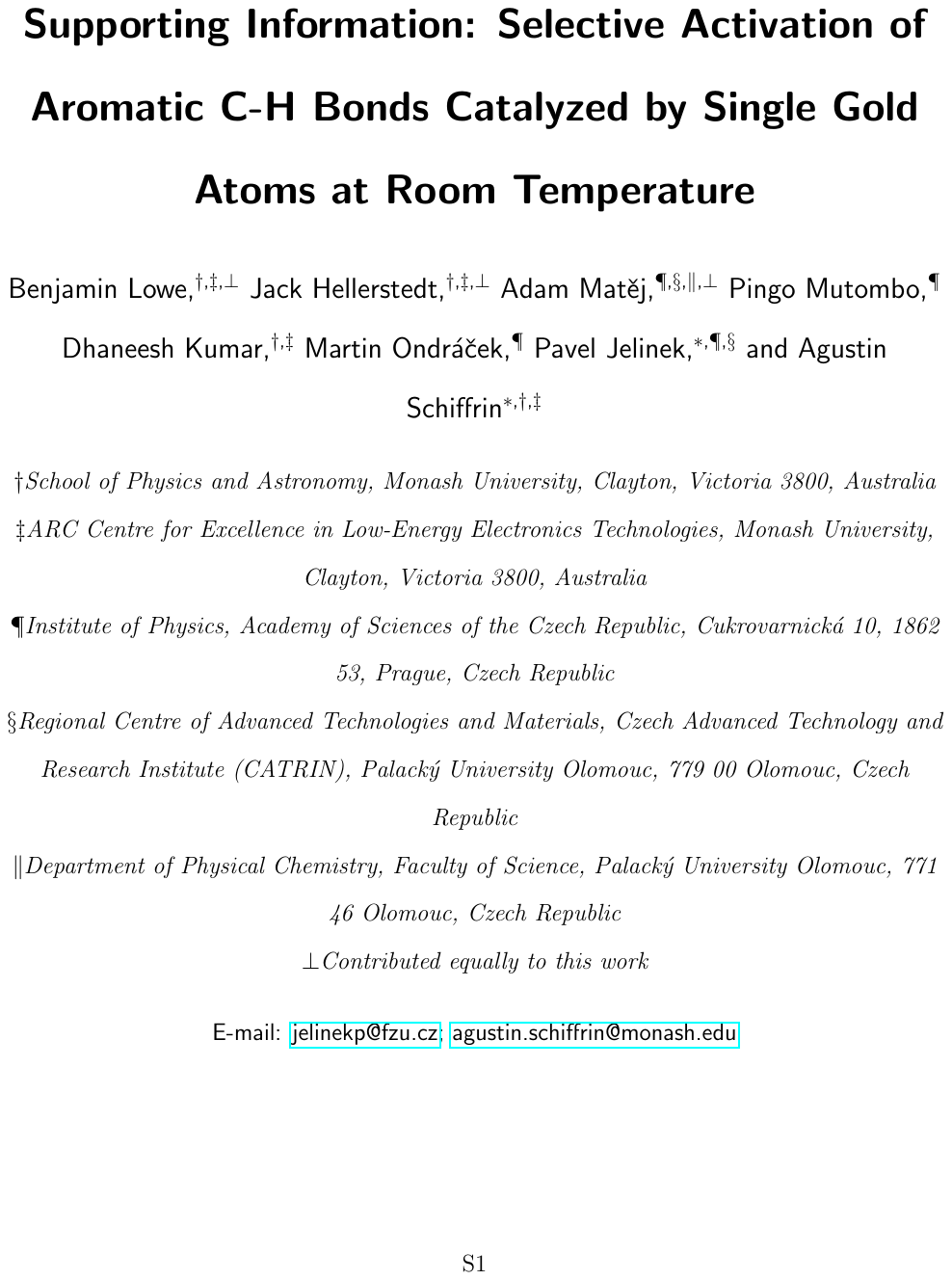}
\end{document}